\documentclass[11pt,psfig,epsfig,preprint]{aastex}
\usepackage{epsfig,psfig}

\begin{document}

\singlespace

\title{Status of CMB Polarization Measurements\\
       from DASI and Other Experiments}

\author{
J.\ E.\ Carlstrom\altaffilmark{1,2,3,4},
J.\ Kovac\altaffilmark{1,3,4},
E.\ M.\ Leitch\altaffilmark{1,2}
and C.\ Pryke\altaffilmark{1,2,4}
}
\altaffiltext{1}{Center for Cosmological Physics, University of Chicago}
\altaffiltext{2}{Department of Astronomy \& Astrophysics, University of Chicago}
\altaffiltext{3}{Department of Physics, University of Chicago}
\altaffiltext{4}{Enrico Fermi Institute, University of Chicago}


\def\tt#1{${\it T}${#1}}
\def\te#1{${\it TE}${#1}}
\def\emode#1{${\it E}$-mode{#1}}
\def\bmode#1{${\it B}$-mode{#1}}
\def\lss#1{large-scale structure#1}
\def\wmap#1{WMAP{#1}}
\def\inflation#1{inflation{#1}}
\def\ss#1{{\it #1}}


\def\citenop#1{\citeauthor{#1} \citeyear{#1}}

\def\fig#1{Figure~\ref{#1}}
\def\sec#1{\S~\ref{#1}}

\def\comment#1{{\bf #1}}

\begin{abstract}
We review the current status and future plans for polarization measurements 
of the cosmic microwave background radiation, as well as the cosmology
these measurements will address.
After a long period of increasingly sensitive upper
limits, the DASI experiment has detected the \emode{} polarization
and both the DASI and \wmap{} experiments have detected the \te{}
correlation. These detections provide confirmation of the 
standard model of adiabatic primordial density fluctuations
consistent with \inflation{ary} models. The \wmap{} \te{} correlation
on large angular scales provides direct evidence of significant reionization
at higher redshifts than had previously been supposed.
These detections
mark the beginning of a new era in CMB measurements
and the rich cosmology that can be gleaned from them. 

\end{abstract}

\section{Introduction}

Tremendous progress has been made in characterizing the angular power
spectrum of the CMB temperature fluctuations over the last decade. As
discussed extensively at the workshop, the temperature anisotropy
spectrum is now well-determined from the largest angular scales
through the first two acoustic peaks and into the third at $\ell \sim
700$. Data on smaller angular scales, in particular from the CBI and
ACBAR experiments, have revealed the damping tail of the primordial
spectrum \citep{runyan03,mason02,kuo03}.  The high-$\ell$ data
provide evidence for the onset of secondary anisotropy at $\ell \sim
2000$ that is also detected at $\ell \sim 5000$ with BIMA
\citep{dawson01}, and that presumably arises at least in part from the
integrated Sunyaev-Zel'dovich Effect from galaxy clusters
\citep[e.g.,][]{komatsu02,holder02}.
Future cosmological studies with the CMB will therefore focus on measurement of
the temperature anisotropy at fine angular scales and on
the polarization anisotropy at all
scales. This paper concentrates on the latter.

Measurements of astronomical polarization are in general difficult and
the low level of the CMB polarization signal makes it an especially
challenging target.  Yet the fundamental nature of
the science 
has fueled rapid progress in
experimental efforts to reach unprecedented levels of sensitivity and
control of systematics. The following is an abbreviated list of the
scientific goals of CMB polarization studies, ranked in increasing
order of required instrumental sensitivity:

\begin{enumerate}

\item {Test the theoretical framework for the generation of CMB
anisotropy: Within the context of the standard model, in which peaks
in the CMB angular power spectrum are interpreted as the signature of
acoustic oscillations seeded by nearly scale-free, primordial
adiabatic density fluctuations, current temperature measurements lead to
highly-specific predictions for the shape of the polarization power
spectrum.  Furthermore, it is a firm prediction that density
fluctuations should create only \emode, i.e., curl-free, polarization
patterns on the sky (see \sec{sec:generation}).}

\item {Determine the reionization history of the universe: When the
universe underwent reionization, electrons re-scattered the CMB,
leading to polarization on the large angular scales corresponding to the horizon
size at the epoch of reionization.}

\item {Improve the precision of CMB-derived cosmological parameters:
Polarization accounts for two thirds of the CMB observables
permitting higher precision constraints, and also allows various
parameter degeneracies to be broken.}

\item {Provide a probe of \lss{} to $z = 1100$. Gravitational lensing
of the CMB by intervening structure distorts the polarization pattern
generated at the surface of last scattering, leading to an observable
\bmode, i.e., curl component. The \bmode, although weak, should be
detectable and can be used to infer properties of the \lss{} \citep{hu02}.}

\item {Test \inflation{} by searching for primordial gravitational waves:
Primordial gravity waves will lead to polarization in the CMB
\citep{polnarev85,crittenden93} with both an \emode{} pattern, as for
the scalar density perturbations, and a \bmode{} pattern, due to the
intrinsic polarization of the gravitational waves
\citep{seljak97a,kamionkowski97b,seljak97}.  In \inflation{ary} models,
the amplitude of the \bmode{} polarization from gravity waves is
proportional to the fourth power of the \inflation{ary} energy scale.
While the detection of \bmode{} polarization would provide a critical
test of \inflation, the signal may be so weak as to be
unobservable \citep{lyth97,kinney03}.}

\end{enumerate}

The current state of CMB polarization studies is similar to the state
of temperature measurements a decade ago, when COBE and FIRS had
detected anisotropy \citep{smoot92,ganga93}, but much work remained to
be done to characterize its power spectrum. Currently, DASI has
detected the \emode{} polarization and \te{} correlation on degree
scales \citep{kovac02}, while \wmap{} has detected the \te{}
correlation on large scales \citep{kogut03a,kogut03b}. The degree-scale
polarization and the \te{} correlation are critical tests of the
underlying theoretical framework for the generation of CMB
fluctuations, while the \wmap{} \te{} correlation on very large scales,
i.e., the first several multipoles, provides direct evidence of significant
reionization at higher redshifts than had previously been supposed.

In this contribution, 
we first briefly review the mechanism which
causes observable polarization within the context of adiabatic density
perturbations, lensing by \lss{} and gravity waves. We
next review the history of polarization measurements, the DASI
and \wmap{} detections (particular emphasis is given to
the DASI experiment as the \wmap{} results are 
also discussed in contributions by \cite{wright03} and \cite{kogut03b}), several upcoming 
and future experiments, 
and conclude with a few remarks on the state of the field.

\section{Generation of CMB Polarization}
\label{sec:generation}

Polarization of the CMB arises 
from
Thomson scattering of the background radiation off electrons at the
epoch of decoupling \citep{rees68}.  As scattering can only produce a
net polarization if an electron sees a local quadrupole moment,
polarization is suppressed as long as the photon mean free path is short, and electrons see
a field that is locally uniform.  As recombination proceeds, however,
the mean free path grows rapidly and electrons begin to see radiation
Doppler-shifted by velocity fields within the plasma.

\begin{figure}[t]
\begin{center}
\epsfig{file=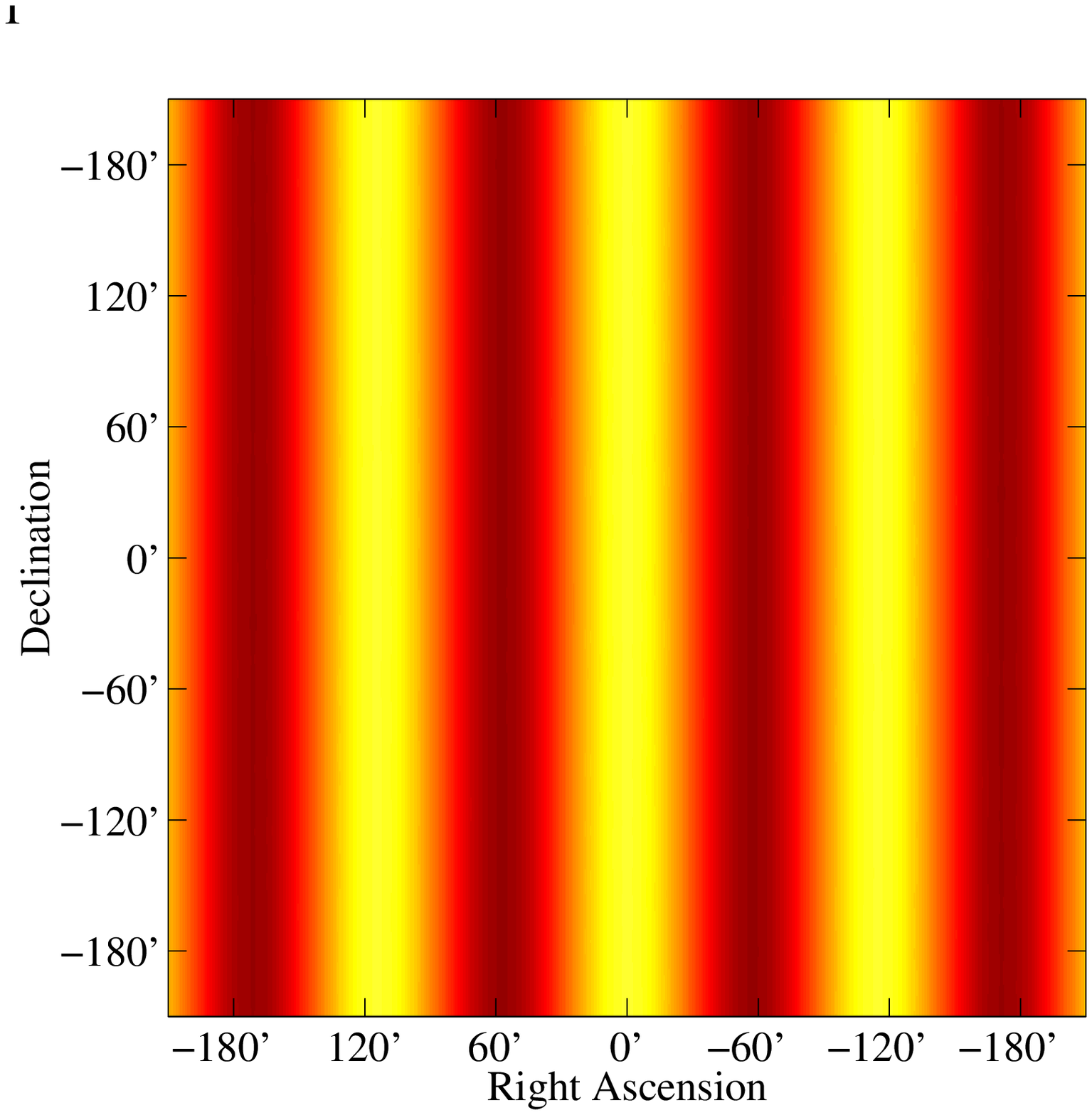,height=2.15in}
\epsfig{file=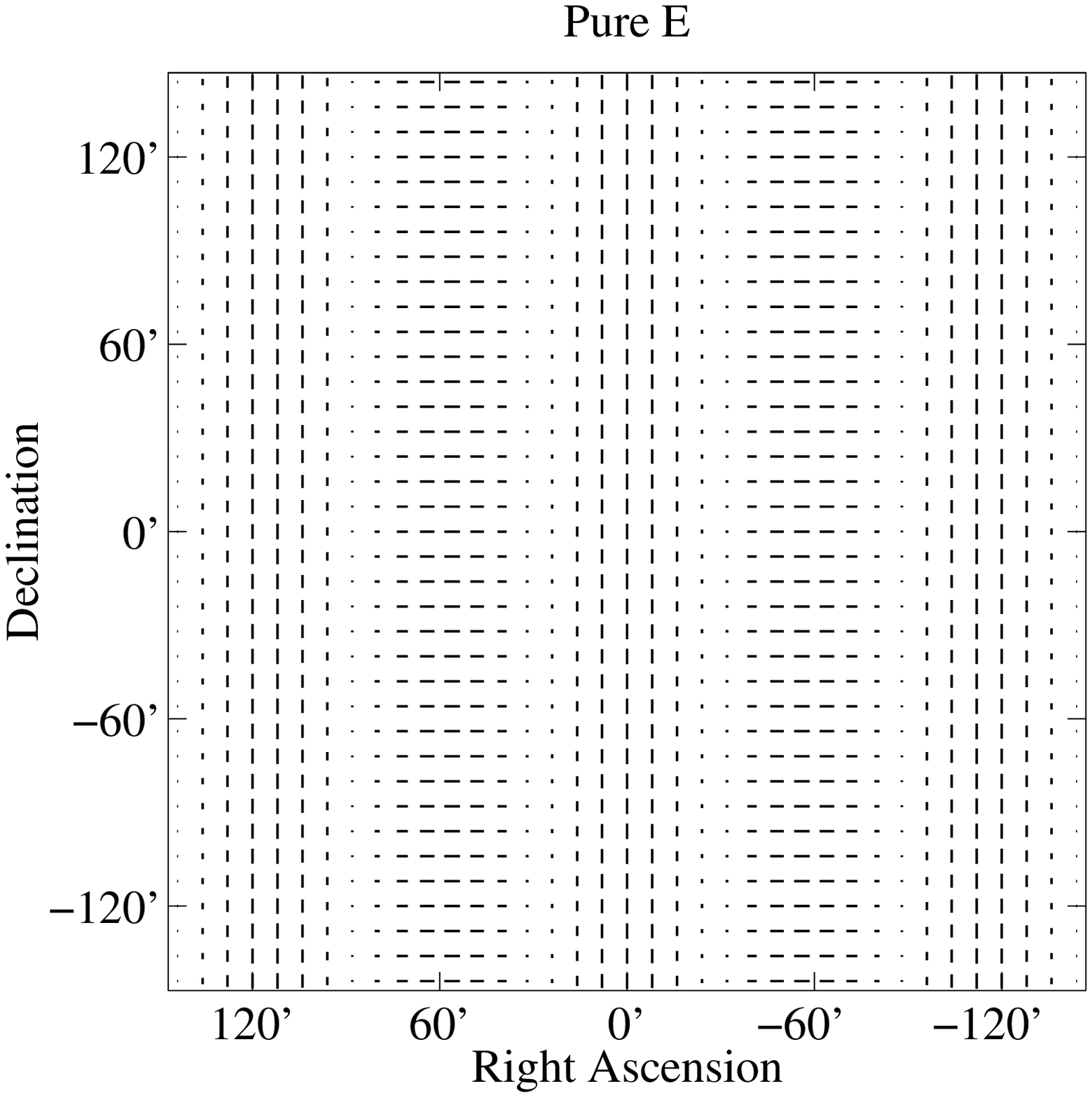,height=2.15in,bbllx=116,bblly=165,bburx=513,bbury=637,clip=true}
\epsfig{file=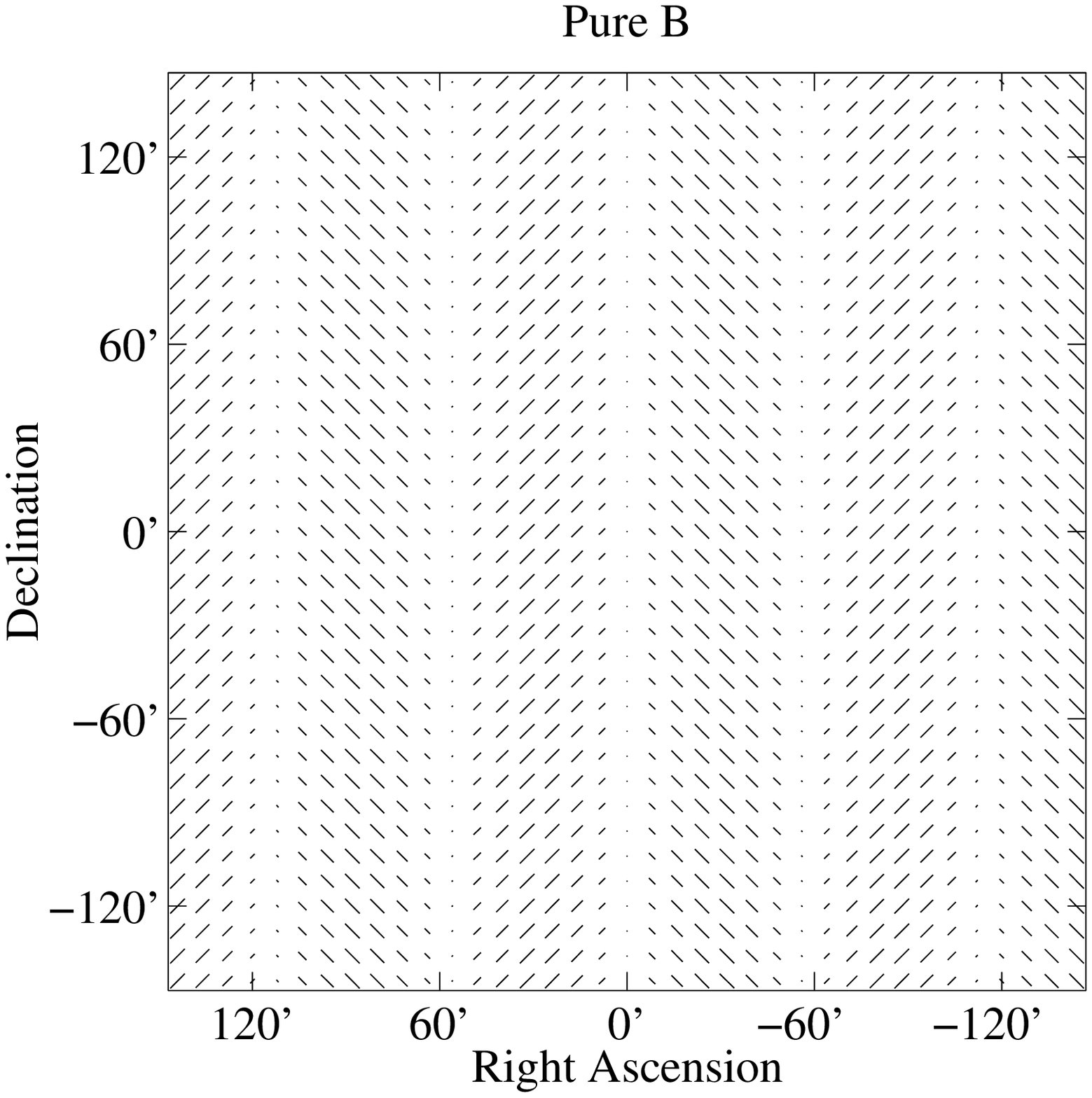,height=2.15in,bbllx=116,bblly=165,bburx=513,bbury=637,clip=true}
\caption{${\it E}$ and \bmode{} polarization patterns.  (left panel) A
representative Fourier mode of a density perturbation. (middle)
\emode{} polarization pattern resulting from Thomson scattering of
this mode (growing amplitude). (right) \bmode{} polarization pattern.}
\label{fig:EBpatterns}
\end{center}
\end{figure}

To illustrate this effect, consider a single spatial mode of the
density field, i.e., a standing acoustic wave; as the amplitude of the
wave oscillates, the photon-baryon fluid will compress and expand. If
the mode is growing, the fluid will move toward density crests. In
this case, the radiation field seen by electrons near a crest is
Doppler-boosted perpendicular to the crest, with no boost seen
parallel to the crest, leading to scattered light whose polarization
is aligned with the crest. Likewise in the troughs, the radiation is
reduced in directions perpendicular to the trough, but uniform along
it, and the resulting polarization will be perpendicular to the
trough, i.e., the sense of the polarization flips between crest and
trough (see \fig{fig:EBpatterns}).  If the amplitude of the 
density mode is decreasing,
the sense of the polarization is reversed, but in general, scattering
leads to polarization aligned either perpendicular or parallel to the
wave vector, a curl-free pattern we refer to as \emode{} polarization
\citep[see also][]{hu_w97,zaldarriaga97}.  The first detection of
\emode{s} in the CMB by the DASI experiment is discussed in
\sec{sec:detections}.

\begin{figure}[t]
\begin{center}
\vskip-0.0in
\epsfxsize=0.6\textwidth
\epsfbox{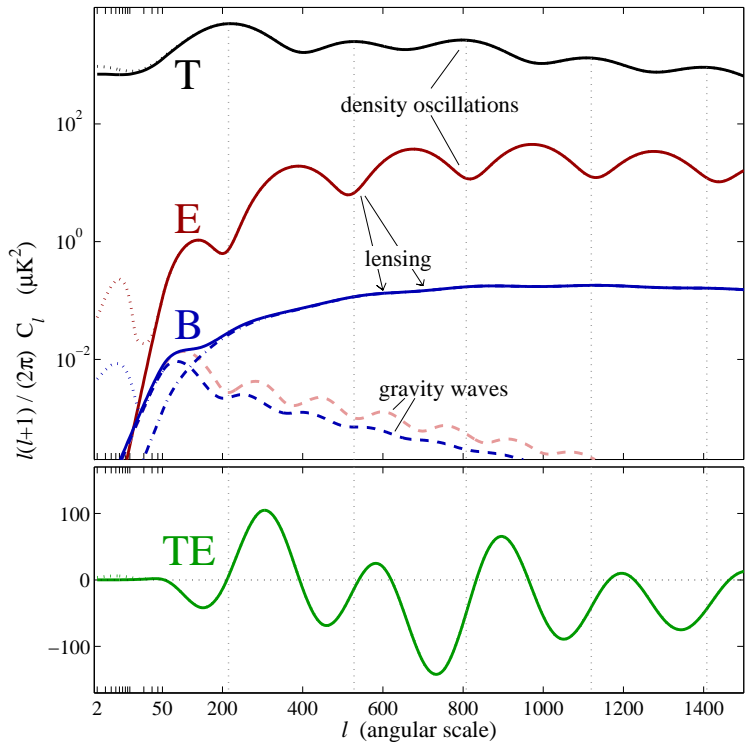}
\parbox{\textwidth}{\caption{\small \label{fig:model-spectra}
Predicted power spectra for the ``standard'' model.  Top to bottom:
(\ss{T}) Temperature, (\ss{E}) \emode{} polarization,
(\ss{B}) \bmode{} polarization, and (\ss{TE})
Temperature-\emode{} cross correlation spectra.
The modification of each spectrum resulting from reionization of the
magnitude observed by \wmap{} is shown by the dotted lines.
Gravitational waves at a level allowed by current data introduce contributions to the
\emode{} and \bmode{} spectra shown by the dashed lines.
The dot-dash line shows the contribution to the \bmode{} signal resulting from
lensed \emode{s}. \citep[Spectra calculated using CMBFAST,][]{seljak96}.}}
\end{center}
\end{figure}

We have so far considered a time snapshot of a single density mode; if
we now consider the dynamic evolution of that mode, we see that
because \emode{} polarization arises from velocities, the amplitude of
the polarization will fall to a minimum at the compression or
expansion maxima of the density mode, when the velocity drops to zero.
Likewise the amplitude of the polarization will be highest at the
density nulls, when the fluid velocity reaches a maximum.  As a
result, the power spectra of the temperature and polarization fields
exhibit peaks which are
approximately a half-cycle out of phase (see \fig{fig:model-spectra}).  The \te{}
correlation spectrum quantifies the complex relation between these
fields, with the sign of the correlation depending on whether the 
amplitude of the mode
was increasing or decreasing at the time of decoupling.

When the universe reionized, late-time Thompson scattering of CMB photons reduced
power in the anisotropy at small scales but regenerated a polarized signal on scales
comparable to the horizon at the reionization epoch, leading to a
significant low-$\ell$ bump in the \emode{} and \te{} correlation
spectra.  Such a signal has now been seen in the \te{} spectrum by the
\wmap{} experiment (see \sec{sec:detections}).

Any polarization field can be decomposed into curl-free
(\emode{}) components, and pure curl components, called \bmode{s} by
analogy with electric and magnetic fields. The $B$-type harmonic modes
exhibit linear polarization at $\pm 45^{\circ}$ to the direction of 
modulation (\fig{fig:EBpatterns}).  Such a pattern
cannot be produced by density modes, and the presence of
\bmode{s} in the CMB would be the distinctive signature of
gravitational effects.  
A stochastic background of primordial gravity waves generated by \inflation{}
would directly source \bmode{s} in the CMB.
Because such waves decay after entering the horizon, 
the spectrum of the resulting \bmode{} signal is expected to peak at large angular scales,
with an amplitude that is tied to the (highly uncertain) \inflation{ary} energy scale.
Within the next decade this signal may be detected yielding exciting
direct evidence for \inflation{}, or may be shown to be unobservably small.
On small angular scales, \bmode{s} are confidently expected from
gravitational lensing of the \emode{} signal by \lss{} (see
\fig{fig:model-spectra}), at a level that should be detectable by the
next generation of dedicated polarimeters (see \sec{sec:upcoming});

\section{Polarization  Measurements}

Prior to the DASI polarization results discussed below
\citep[see][]{kovac02}, only upper limits had been placed on the level
of CMB polarization, a measure both of the demanding instrumental
sensitivity and attention to sources of systematic
uncertainty necessitated by the weakness of the expected signal
\citep[see][for a review of CMB polarization measurements]{staggs99}.
We review these limits below, followed by a discussion of the recent
detections of polarization by the DASI and \wmap{} experiments.

\begin{figure}[t]
\begin{center}
\vskip-0.0in
\epsfxsize=0.6\textwidth
\epsfbox{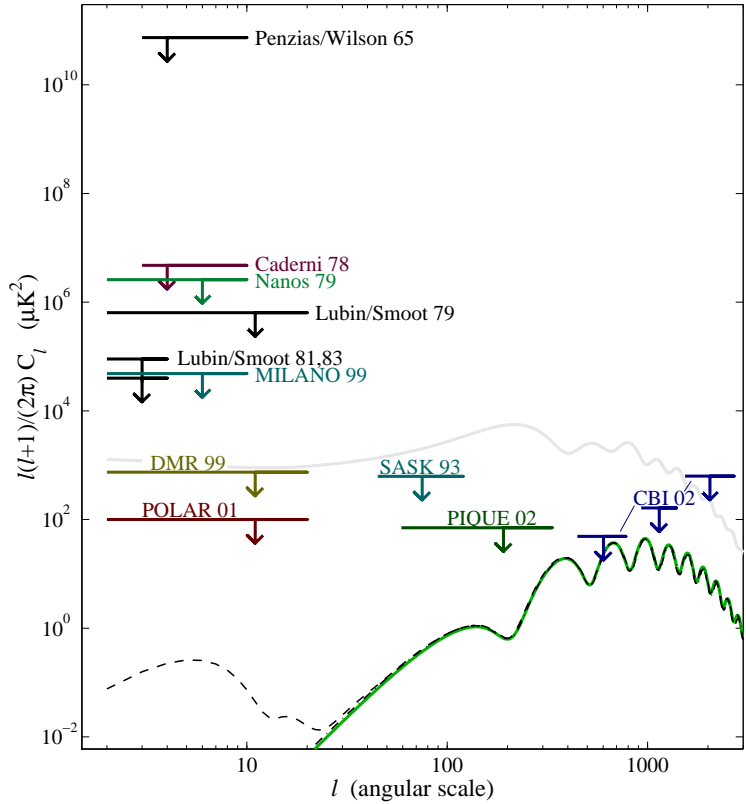}
\parbox{\textwidth}{\caption{\small \label{fig:exp-limits}
Experimental limits to the polarization of the CMB anisotropy prior to
the 2002 DASI detection. See text for references.}}
\end{center}
\end{figure}

\subsection{Early Limits}
\label{sec:limits}

The first constraint on the degree of polarization of the CMB was
placed in 1965 by its co-discoverers Penzias and Wilson, who stated
that the new radiation they had detected was isotropic and unpolarized
within the limits of their observations \citep{penzias65}.  Over the
next 20 years, dedicated polarimeters were used to set much more
stringent upper limits on angular scales of several degrees and larger
\citep[][see also
\citenop{sironi97}]{caderni78,nanos79,lubin79,lubin81,lubin83}, with
the best upper limits for the \emode{} and \bmode{} polarizations
being 10$~\mu$K at 95\% confidence for the multipole range $2 \le \ell
\le 20$, from the POLAR
experiment \citep{keating01}. The POLAR experiment was reconfigured
to the COMPASS experiment at intermediate scales \citep{farese03}. 

An analysis of data from the Saskatoon experiment \citep{wollack93}
set the first upper limit on somewhat smaller angular scales
($25~\mu$K at 95\% confidence for $\ell \sim 75$); this result is also
noteworthy for being the first that was below the level of the CMB
temperature anisotropy. The best limit on similar angular
scales was set by the PIQUE experiment \citep{hedman02} --- a
95\% confidence upper limit of $8.4~\mu$K to the \emode{} signal,
assuming no \bmode{} polarization.  Analysis of CBI data
set upper limits similar to the PIQUE
result, but on somewhat smaller scales \citep{cartwright_thesis}.  An attempt was also made
to search for the \te{} correlation using the PIQUE polarization and
Saskatoon temperature data \citep{costa02}.

Polarization measurements have also been pursued on arcminute scales,
resulting in several upper limits
\citep[e.g.,][]{partridge97,subrahmanyan00}.  However, at these
angular scales, corresponding to multipoles $\sim 5000$, the level of
the primary CMB anisotropy is strongly damped and secondary effects
due to interactions with \lss{} in the universe are expected to
dominate \citep{hudodelson02}.

\subsection{Detections}
\label{sec:detections}

\begin{figure}[t]
\begin{center}
\vskip-0.0in
\epsfxsize=0.51\textwidth
\epsfbox{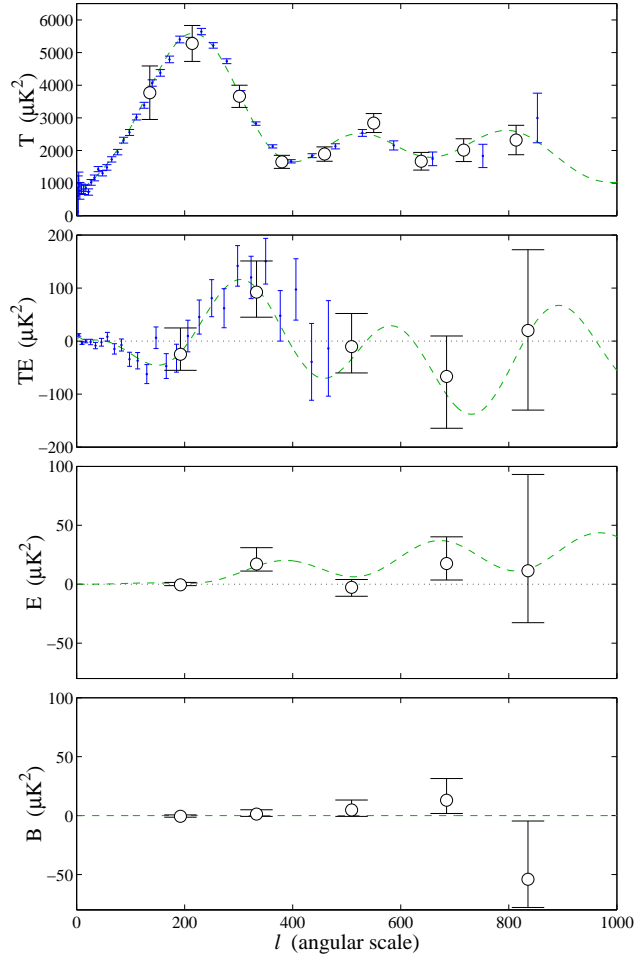}
\parbox{\textwidth}{
  \caption{\small \label{fig:DASI-WMAP} Recent detections of CMB
    polarization with DASI and \wmap.  Panels show experimental
    bandpowers of, in order from top to bottom, temperature \tt, \te{}
    correlation, \emode, and \bmode{} polarization.  DASI points are
    shown as open circles, \wmap{} as small closed symbols.}}
\end{center}
\end{figure}

As can be seen in \fig{fig:exp-limits}, high-resolution experiments
have steadily converged on the sensitivity required to see the
\emode{} signature expected under the standard model.  With the
results from the DASI experiment, reported in \citep[][see also
\citenop{kovac_thesis} for details]{kovac02}, the goal of direct
detection of the \emode{s} has now been achieved, with the ancillary
detection of the \te{} correlation.  More recently, the \wmap{}
experiment has reported the detection of the \te{} correlation on
large-scales.  These results are summarized in \fig{fig:DASI-WMAP}.

\begin{figure}[t]
\begin{center}
\vskip-0.0in
\epsfxsize=0.4\textwidth
\epsfbox{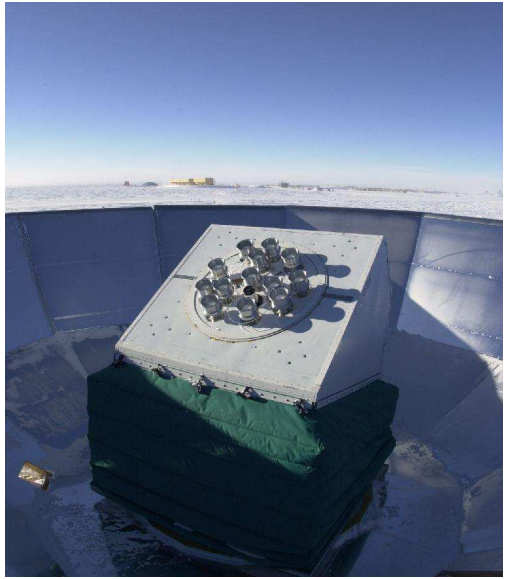}
\parbox{\textwidth}{\caption{\small \label{fig:DASI-photo} DASI in its
ground shield, showing the feeds and movable faceplate. The
Amundsen-Scott South Pole Station can be seen in the background.}}
\end{center}
\end{figure}

\begin{figure}[tbp]
\begin{center}
\epsfig{file=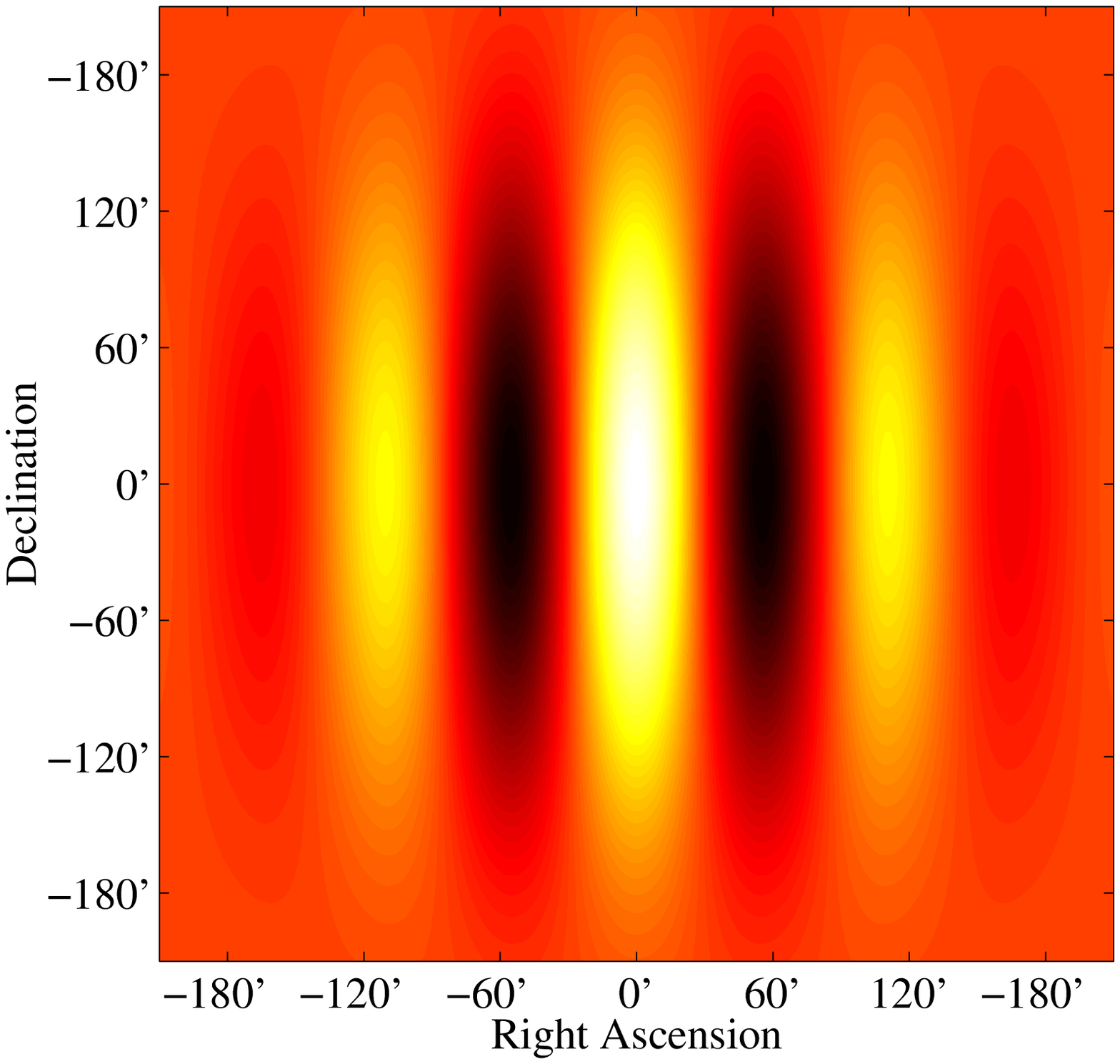,height=2.0in}
\epsfig{file=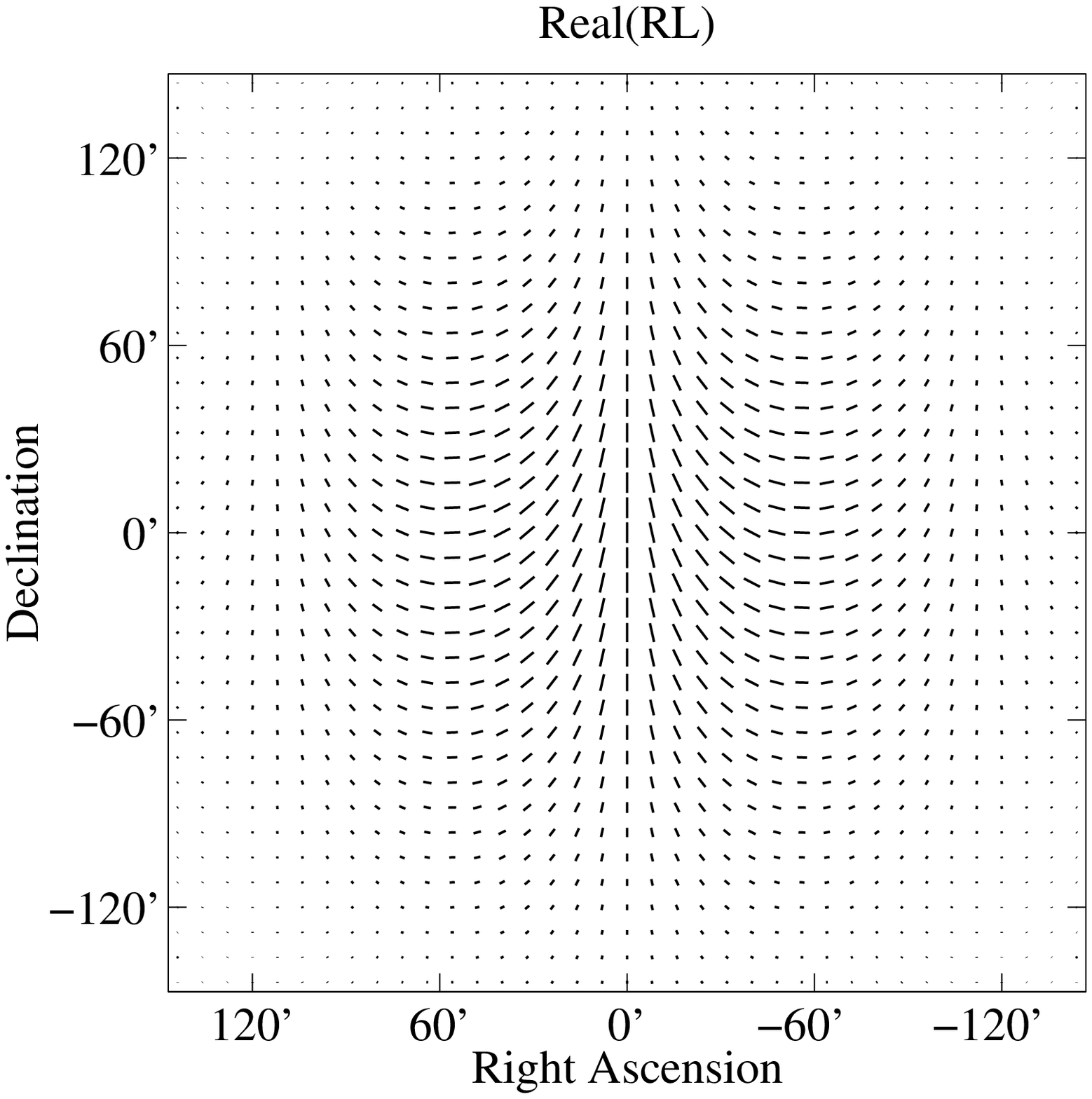,height=2.15in,bbllx=116,bblly=165,bburx=513,bbury=637,clip=true}
\epsfig{file=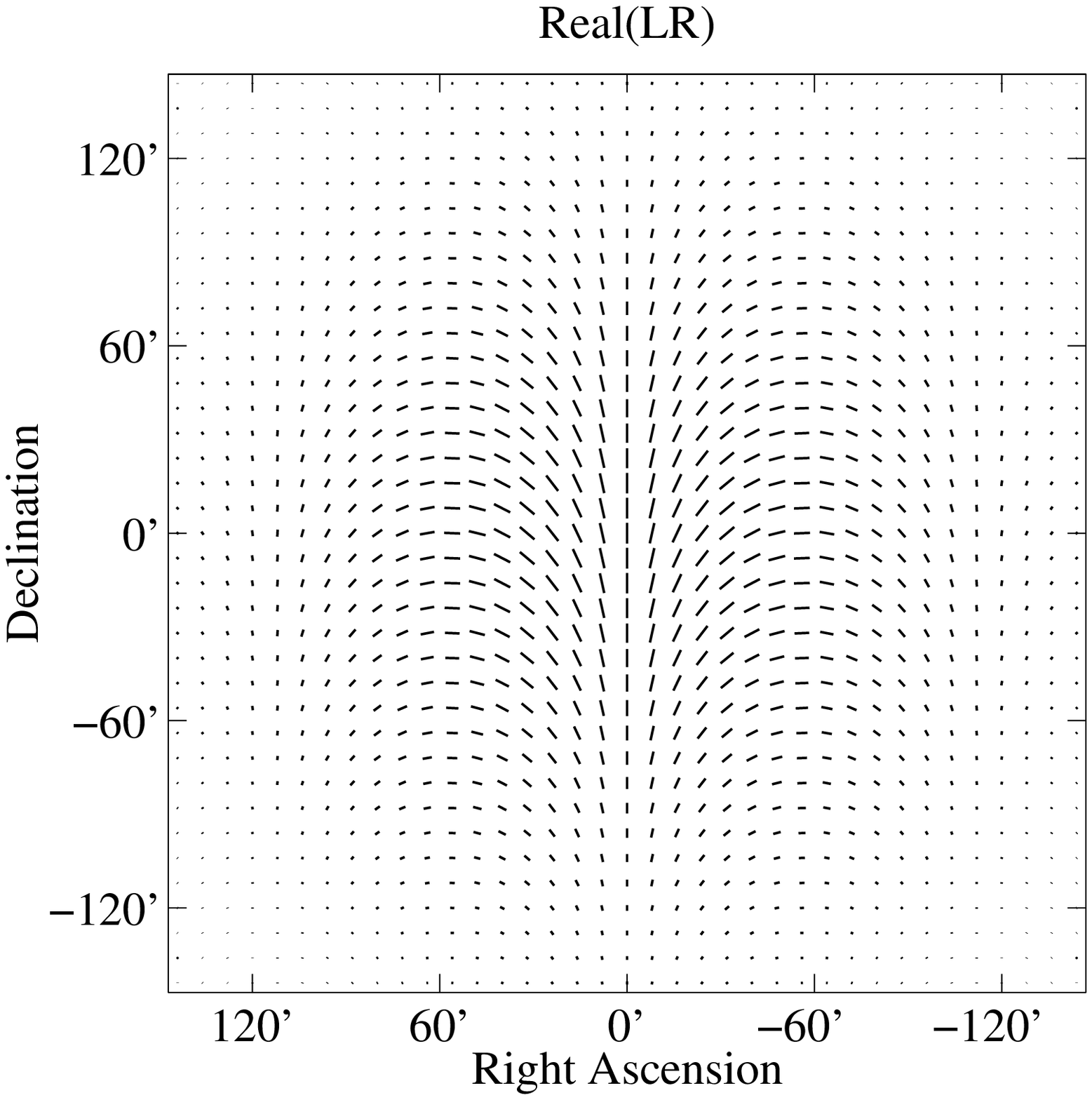,height=2.15in,bbllx=116,bblly=165,bburx=513,bbury=637,clip=true}
\caption{Effective response of a single baseline of an interferometer,
for total intensity (left panel), and cross-polarized baselines (right
panels) (see \sec{sec:detections}).  It can be seen that the intensity
response differs from a pure matched Fourier filter only by the taper
of the primary beam.}
\label{fig:int-response}
\end{center}
\end{figure}

The DASI experiment exploits the unique properties of interferometry
directly to measure Fourier components of the CMB anisotropy with
tight control of systematics.  As interferometers are correlating
devices, they are largely insensitive to sources of incoherent noise,
while providing an effective response on the sky that is a
near-perfect Fourier filter (see \fig{fig:int-response}). DASI's
elements are mounted on a faceplate that can be rotated to sample
different Fourier modes, or to sample the same modes with independent
antennas.  DASI was used successfully to measure the temperature
anisotropy spectrum during the austral 2000 winter; the design and
calibration of the instrument, the power spectrum, and the resulting
cosmological constraints are reported in \cite{leitch02a},
\cite{halverson02} and \cite{pryke02}, respectively. For additional
details see \cite{halverson_thesis}.

\begin{figure}[tbp]
\begin{center}
\epsfig{file=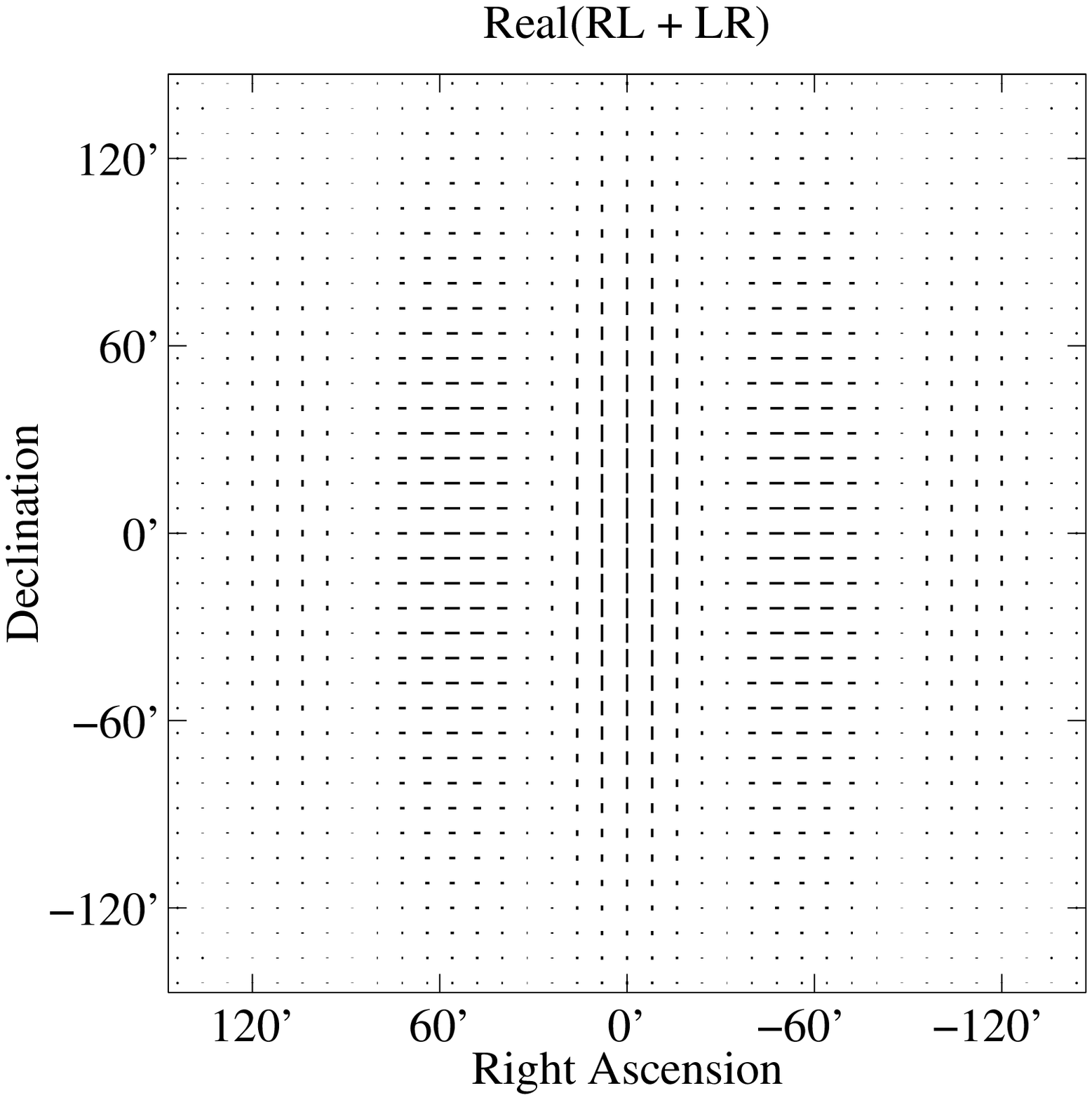,height=2.15in,bbllx=116,bblly=165,bburx=513,bbury=637,clip=true}
\epsfig{file=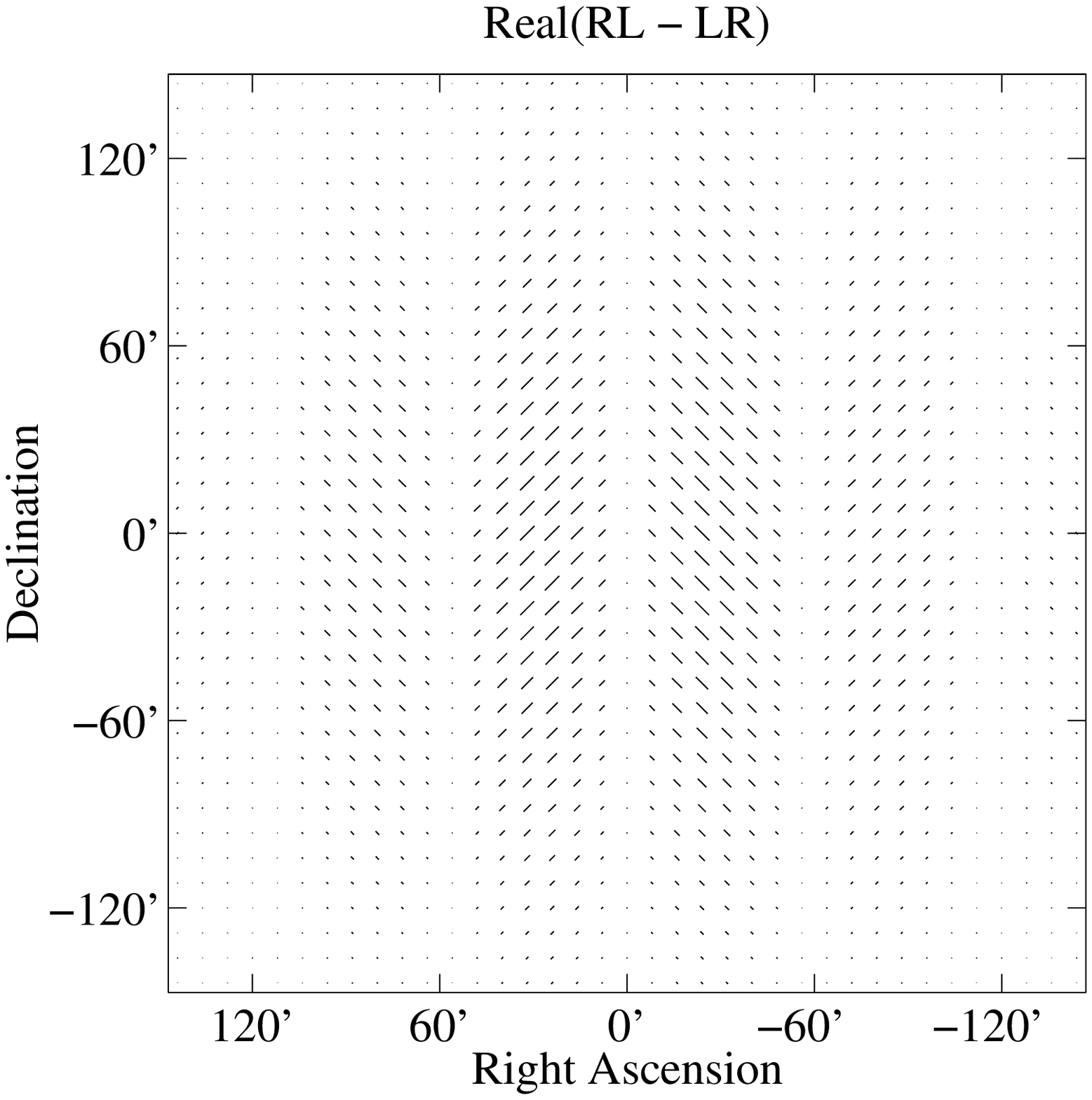,height=2.15in,bbllx=116,bblly=165,bburx=513,bbury=637,clip=true}
\caption{Linear combinations of the DASI cross-polar baseline
responses shown in \fig{fig:int-response}.  It can be seen by
comparison with \fig{fig:EBpatterns} that combinations of
$\ss{RL}\pm\ss{LR}$ differ from pure \emode{} and \bmode{} responses
only by the taper of the primary beam, which results in a tiny and
easily-characterized leakage between responses to the two modes \citep{kovac_thesis}.}
\label{fig:int-E-B}
\end{center}
\end{figure}

The reconfiguration and operation of the instrument for polarization
observations are described in detail in \cite{leitch02b}.  During the
2000---2001 austral summer, broadband achromatic polarizers were
installed in each of the 26--36 GHz receivers, and a large reflecting screen
was erected to reduce the sensitivity to contamination from the
ground.  By mechanically switching the polarizers for each receiver to
pass left (\ss{L}) or right (\ss{R})-circularly polarized light, each
baseline (a single pair of antennas whose signals are correlated) can
sample the full complement of Stokes parameters, where co-polar
(\ss{RR} \& \ss{LL}) states are sensitive to the total intensity, and
cross-polar (\ss{RL} \& \ss{LR}) states are sensitive to linear
polarization (see \fig{fig:int-response}).  As can be seen in
\fig{fig:int-E-B}, simple combinations of the cross-polar data
produce nearly pure \emode{} and \bmode{} responses on the sky.

Since the austral summer of 2001, DASI has observed two regions of
sky, with a field of view of approximately $3\fdg4$.  A host of
consistency tests on the data demonstrate that the combination of
shielding, field-differencing and careful characterization of the
instrumental polarization has reduced any common-mode residuals to
well below the level required to detect CMB polarization, resulting in
one of the deepest integrations ever achieved on the CMB.  The
effective noise on the DASI temperature map of the CMB in these fields
is approximately $2.7\mu$K, and has been demonstrated to
integrate down with the square-root of time from timescales of 
seconds to years.  

The DASI data have yielded a direct detection of \emode{} polarization;
parameterizing the power spectrum as a single shaped bandpower over
the full $\ell$-range, we find that \emode{} polarization is detected
at $4.9\sigma$ significance, with likelihood ratio tests demonstrating
that the data strongly prefer the concordance model shape to flat or
power law alternatives.  Results with the various power spectra
parameterized in five bands are shown in \fig{fig:DASI-WMAP}.  The
shape and amplitude of the \emode{} spectrum are consistent with
predictions from the current best-fit model to the temperature data.

\begin{figure}[t]
\begin{center}
\vskip-0.0in
\epsfxsize=0.5\textwidth
\epsfbox{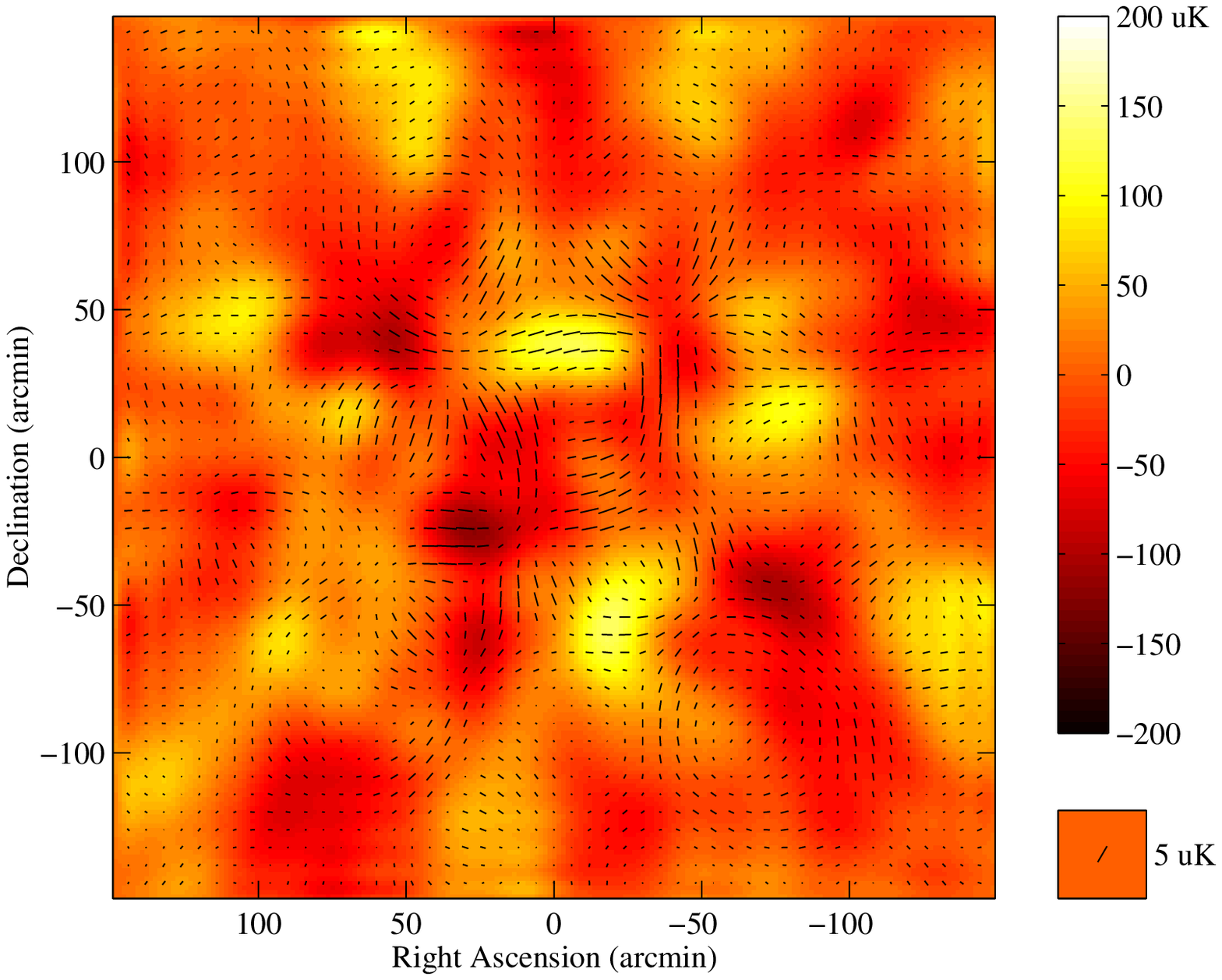}
\parbox{\textwidth}{\caption{\small \label{fig:DASI-temap} Difference
    map of the two fields observed in the DASI polarization
    experiment.  Grayscale is the temperature map, showing high
    signal-to-noise detection of CMB structure.  Noise in this map is
    approximately $2.7\mu$K.  Vector overlay is the \emode{}
    polarization map constructed from the high signal-to-noise modes
    in the data \citep{kovac_thesis}.}}
\end{center}
\end{figure}

The strategy of very deep integration on a small region of sky
yielded not only a statistical detection of polarization with DASI, but a data
set containing many high signal-to-noise polarization modes.  The
temperature map, and the polarization map constructed from these high
signal-to-noise modes, are shown in \fig{fig:DASI-temap}.

\begin{figure}[thb]
\begin{center}
\vskip-0.0in
\epsfxsize=0.6\textwidth
\epsfbox{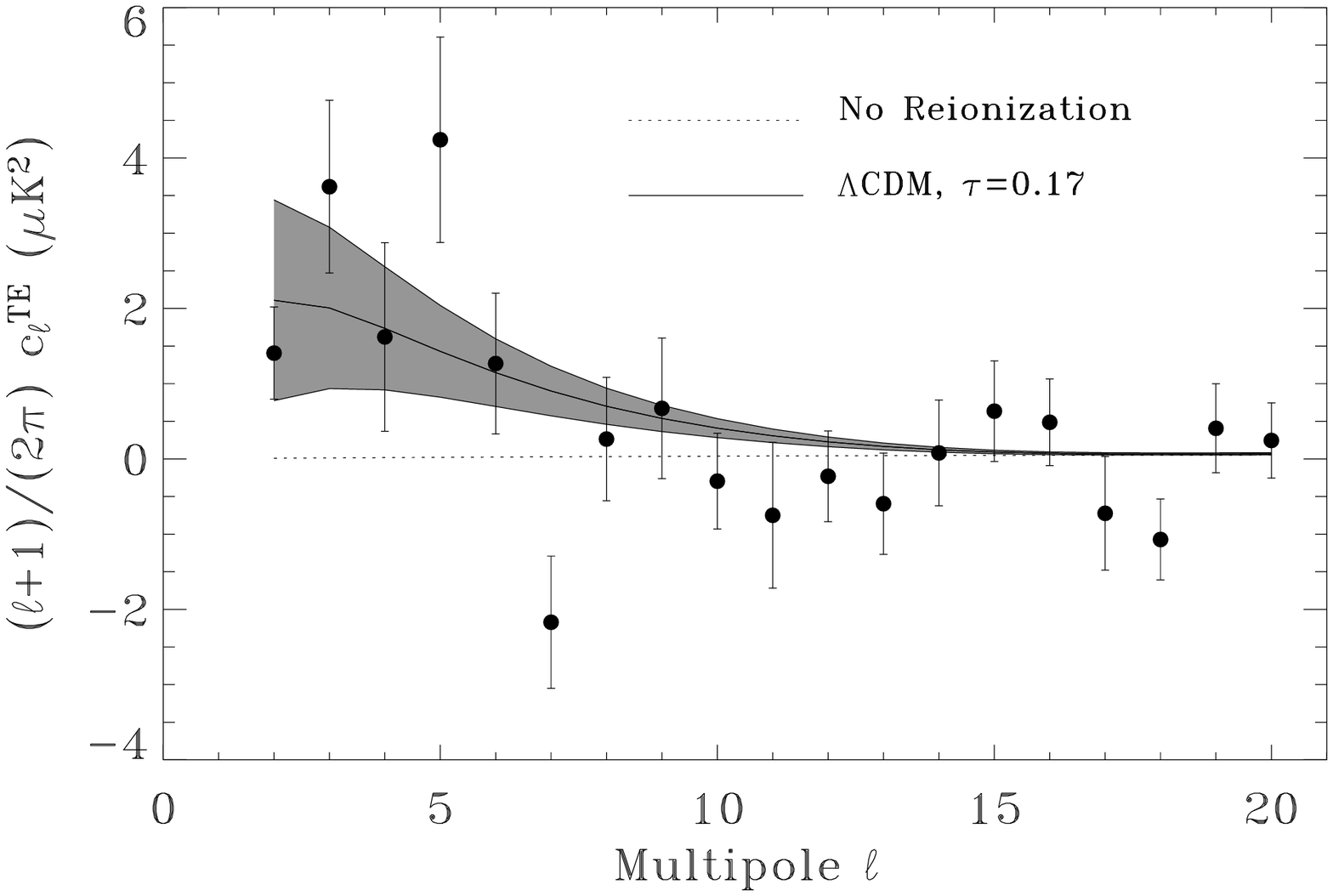}
\parbox{\textwidth}{\caption{\small \label{fig:WMAP-TE} 
The low $l$ portion of the polarization cross-power spectra $c_\ell^{TE}$ 
for the \wmap{} one-year data \citep{kogut03b}.
The excess power in the first few multipoles indicates significant reionization
at high redshift.
(Note that
$(\ell+1)/2\pi~ c_l^{TE}$ is plotted rather than
$\ell(\ell+1)/2\pi~ c_l^{TE}$.)
}}
\end{center}
\end{figure}

Both DASI and \wmap (see workshop
contributions by \cite{wright03} and \cite{kogut03b}) have also
detected the distinctive \te{} correlation of the CMB, which is a
simple consequence of the fact that the fluid velocities which lead to
polarization (see \sec{sec:generation}) are sourced by the same
density fluctuations which lead to temperature anisotropy (second
panel of \fig{fig:DASI-WMAP}).  The power of full-sky mapping
experiments like \wmap{} is evident on the largest scales, where the
DASI measurement of the \te{} correlation is limited by sample
variance on the temperature signal.  The first-year data from \wmap{}
have resulted in confirmation of the predicted
large-scale \te{} correlation, and extraordinary
direct evidence of significant reionization
at higher redshifts than had previously been supposed
(see \fig{fig:WMAP-TE}).

\section{Ongoing and Upcoming Experiments}
\label{sec:upcoming}

\subsection{Ongoing Experiments}

In addition to DASI and \wmap{}, both of which are 
still collecting data, there are several ongoing CMB polarization experiments
that have recently obtained data. Like DASI and \wmap{} they
can be classified as second-generation experiments and are
likely to result in detections. As second generation
experiments they have benefited from the lessons learned from
the pioneering experiments discussed in \S\ref{sec:limits}.

\subsubsection{ground-based}

Ongoing ground-based experiments include DASI (\S\ref{sec:detections}), the
Cosmic Background Imager (CBI, \cite{padin02}) and
the Cosmic Anisotropy Polarization Mapper (CAPMAP, \cite{barkats03}).

The CBI is a companion instrument to DASI which uses low-noise 26--36 GHz HEMT
amplifiers \citep{pospieszalski00b} and operates from the Atacama plateau in Chile. The
thirteen 0.9 meter on-axis Cassegrain interferometer elements are
mounted on a common Alt-Az platform which also can be rotated along
the line of sight. In early 2003, the receivers were fitted with the
same selectable achromatic polarizers developed for DASI
\citep{leitch02b}. Like DASI, each receiver is sensitive to only one
circular polarization state and therefore the interferometer is a
factor of four slower than optimum. The benefit of this scheme,
however, is that the existing 10 GHz bandwidth, 13-element, correlator
\citep{padin00} does not need to be expanded. The polarization
sensitivity of the CBI expressed in units of temperature is
similar to DASI's but at three times higher angular resolution, i.e.,
for similar integration time, the CBI should achieve roughly the same
polarization sensitivity as DASI in units of ${\ell(\ell+1)C_\ell}$,
but at three times higher $\ell$. Scaling from 
the DASI results shown in Fig.~\ref{fig:DASI-WMAP}, we see
that the CBI $\ell$-range and sensitivity should be well matched
for characterizing the \emode{} spectrum.

The CAPMAP experiment, run by Suzanne Staggs and collaborators, uses
 the radiometers and techniques developed for the PIQUE
experiment (\S\ref{sec:limits}; \cite{hedman02}) on the Bell
Laboratory 7-meter off-axis telescope. They operated CAPMAP in 2003
February through April with a limited set of four W-band (14 GHz
bandwidth at 90 GHz) polarization-sensitive correlation receivers;
they expect to use the full set of sixteen 90 GHz radiometers and four
40 GHz radiometers starting in November 2003. The
large telescope enables them to achieve $4'$ resolution -- well
suited for measuring the polarization power spectrum from where it peaks
near $\ell \sim 1000$ to higher $\ell$. The weather in New Jersey,
however, limits observing to the winter months.  They plan to observe
a 1-degree patch at the North Celestial Pole using azimuth scans and
expect to measure the \emode{} spectrum in two winter seasons
\citep{barkats03}. While the sensitivity of the individual CAPMAP
channels at $\sim 1 mK s^{1/2}$ is roughly a factor of two to three
times worse than that predicted for ground-based bolometer
polarization receivers, CAPMAP is using a technique for which it has
been shown that systematics can be controlled and very long
integrations times are feasible.

\subsubsection{balloon-borne}

The current generation of balloon-borne experiments include 
Boomerang \citep{montroy03} and MAXIPOL \citep{johnson03}. 
The Archeops balloon-borne CMB experiment has also 
made high frequency measurements of Galactic dust polarization 
over the angular scales relevant to CMB studies \citep{benoit03}. 
Boomerang made a successful Antarctic long-duration balloon (LDB) flight in 
2003 January and collected 11.7 days of data. MAXIPOL had
a successful 26 hour flight in 2003 May from Fort Sumner, New Mexico.

The focal plane of Boomerang was reconfigured for its 2003 LDB
flight with polarization-sensitive bolometer (PSB) elements. 
PSBs use a micro-mesh design like the spider-web bolometer
design, except that the mesh is in a square grid. The grid
is metalized in only one direction to absorb only one linear
polarization state of the incident radiation. A pair of PSBs
is formed by using two orthogonal elements separated by $60\mu m$.
A pair of PSBs can then used with a single feed with the advantage
that the pair share all the same optics, sidelobes, etc. Boomerang
uses four horns to feed four PSB pairs operating at 145 GHz. In addition, there
are four horns which each feed a bolometers 
at 245 GHz and 345 GHz to measure foreground emission
from Galactic dust. For these channels, the polarization has been selected
by a wire grid located in front of the horn.  At 145 GHz, the resolution
is $9.5'$ and the sensitivity of order 160 $\mu K_{CMB} s^{1/2}$.
They observed a shallow region of 1161 square degrees and
a deep region of 123 square degrees.

The MAXIMA experiment \citep{hanany00} was converted to the
polarimeter MAXIPOL by adding a rotating half-wave plate and fixed
polarizing grids in front of the feed horns \citep{johnson03}. The
half-wave plate spinning at two Hz causes the polarization
sensitivity to be modulated at four Hz. Due to the rotating half-wave
plate, all Stokes parameters
are measured by each detector.  While rotating half-wave
plates have been used often in Galactic submillimeter observations,
e.g., \cite{hildebrand03}, this would be the first successful
implementation of the technique for CMB measurements. The MAXIPOL
focal plane is cooled to 100~$mK$ and by scaling the previous MAXIMA
sensitivities we can expect sensitivities of order 130 $\mu
K_{CMB} s^{1/2}$ for each of the twelve 140 GHz channels. The beam
size is $10'$ and the scan length is 2 degrees.  MAXIPOL also has four
420 GHz channels to measure foreground polarization emission from
Galactic dust.

The Boomerang and MAXIPOL experiments each have sufficient sensitivity to
detect the \emode{} polarization at intermediate angular scales;
assuming the systematics are understood and controlled, the Boomerang 
sensitivity should allow the \emode{} spectrum to be 
characterized beyond what has been reported by DASI.

\subsection{Upcoming Experiments}

\subsubsection{ground-based}

Two new ground-based bolometric array polarimeters are being
developed: BICEP (Background Imaging of Cosmic Extragalactic
Polarization, \cite{keating03b}) and QUEST (Q and U Extra-galactic Survey
Telescope, \cite{church03}). BICEP is scheduled to deploy
to the South Pole for observations starting in Austral Winter 2005.
QUEST is expected to be mounted on the DASI telescope 
(hereafter referred to as QUAD for QUEST and DASI) and to
also start observations in Austral Winter 2005\footnote{The QUEST
radiometer and the optics, including the 2.6 meter Cassegrain
reflector and secondary are funded, but at the time of this
writing the proposal to NSF-OPP to deploy QUEST on 
DASI at the South Pole
is pending.}.

The BICEP and QUAD experiments are in many ways companion
experiments. Both exploit PSB detectors being
developed at JPL and share many team members. They are also
complementary in $\ell$ space, with BICEP targeting
degree angular scales $10<  \ell < 200$ and QUAD targeting
$100 < \ell < 2000$. The sensitivity of each experiment 
should allow the \emode{} spectrum to be well characterized.
Furthermore, QUAD should be able to detect the gravitational lensed
\bmode{} spectrum. And, at degree angular scales, BICEP should be able
to reach the required sensitivity either to detect
the primordial gravitational wave \bmode{} signal,
or to further constrain the \inflation{ary} energy scale,
with the caveats that systematics must be controlled to unprecedented levels ($< 0.1 \mu K$)
and that foregrounds must also be understood at these levels. In any case,
we are likely to learn much from these instruments
about the experimental and foreground challenges that will
need to be met by extremely deep future CMB polarization measurements.

BICEP will consist of an array of 48 horns feeding
PSB detector pairs, half at 100 GHz and half at 150 GHz
with 1.0$^\circ$ and 0.7$^\circ$  resolution, respectively. 
The 150 GHz detectors are expected to have an 
sensitivity of 280 $\mu K_{CMB} s^{1/2}$.  Cooled lenses form the
optical system,
providing a 20 degree field of view. 
The plan is to observe the South Celestial Pole with the
optical boresight offset 5$^\circ$ to 25$^\circ$ from the
zenith and rotating the entire array about the line
of sight roughly once per minute. A novel ferrite Faraday
modulator for each feed will be used to rotate the linear polarization. 

QUAD \citep{church03} consists of
a radiometer similar to BICEP, but used at the
Cassegrain focus of a 2.6-meter on-axis precision telescope. The
secondary will be held by a low-loss foam cone as was done
successfully for COMPASS \citep{farese03} and CBI \citep{padin02}.
The entire telescope will be mounted on the existing DASI mount,
replacing the faceplate on which the DASI receivers
and horns currently reside, as shown in Fig.~\ref{fig:DASI-photo}.
The QUEST focal plane consists of 12 PSB pairs
at 100 GHz and 19 pairs at 150 GHz. The expected
sensitivity is 300 $\mu K_{CMB} s^{1/2}$ at 150 GHz
with a $4'$ beam. The DASI mount is fully steerable and
also can be rotated along the line of sight. The latter
feature will be used along with a rotating half-wave
plate to modulate the polarization response. The 
half-wave plate will be not be synchronously
rotated as for MAXIPOL, but rather set at a fixed
angle for each scan.

The Array for Microwave Background Anisotropy (AMiBA\footnote{see http://amiba.asiaa.sinica.edu.tw/}) is being built for measurements of the fine-scale 
CMB temperature
and polarization anisotropy as well as for observations of the Sunyaev-Zel'dovich Effect. The project is led by the Academia Sinica Institute
of Astronomy and Astrophysics in Taiwan and the array will be deployed
to Mauna Kea, Hawaii. Like the CBI, AMiBA is an interferometric
array mounted on a common platform. 
The specifications
call for 19 elements operating at 90 GHz with full polarization capabilities
and a 20 GHz correlation bandwidth. Two sets of array dishes, 1.2 meter
and 0.3 meter are planned. Initial observations targeting
the \emode{} spectrum with a subset of the array are planned to start
in 2004.

\subsubsection{satellite-based}

The Planck Surveyor \footnote{see
http://astro.esa.int/SA-general/Projects/Planck/} \citep{lamarre03})
is a dedicated CMB satellite scheduled to launch in 2007 to measure
the entire sky in nine frequency bands using coherent, HEMT-amplified
radiometers at 30, 40 and 70 GHz and bolometric detectors at
100, 143, 217, 353, 545 and 857 GHz. While initially designed
primarily for CMB temperature measurements, it has considerable
polarization sensitivity. The HEMT amplifier correlation receivers are
intrinsically polarized and there are 4 at 30 GHz, 6 at 44 GHz and 12
at 70 GHz \citep{mennella03}. The resolution provided by the single 1.5 meter aplanatic
primary ranges from $33'$ to $14'$ across these bands. There are four
unpolarized detectors for each of the bolometer bands, providing
excellent frequency coverage and resolution of $9.2'$ at 100 GHz,
$7.1'$ at 143 GHz, and $5'$ at all the higher frequency bands. The
prime polarization sensitivity is provided by four PSB pairs for each
of the 100, 143, 217 and 353 channels. The polarization sensitivity
for the resulting Planck maps is expected to be of order 4 $\mu K$ per
pixel for the 100 and 143 GHz maps over the entire sky. Comparing
Planck to the expected state of the art for ground and balloon-borne
experiments 5 years from now, is similar to comparing \wmap{} to
existing experiments. The ground-based and balloon-borne experiments
are expected to make deeper polarization maps but over much smaller
regions. Planck's coverage of the entire sky is unique and
valuable and its wide frequency range will be unsurpassed for
understanding foregrounds. Projections for Planck's power
spectrum results can be found in the excellent CMB review
by \cite{hudodelson02}.

Planck is the next generation satellite, both more ambitious and riskier
than \wmap. \wmap{} was designed with emphasis on control of systematics and
calibration even at the expense of sensitivity. There are no
active coolers (or heaters) on \wmap{} and as a result the focal plane
runs warmer than optimum. To achieve high sensitivity Planck's
bolometers require active cooling. \wmap{} used correlation radiometers
with receivers fed by completely separate telescopes pointing
140 degrees apart on the sky, providing simultaneous differencing
on these scales. Planck has a single optical system and
focal plane shared by coherent and
bolometric radiometers. On large scales, it will be difficult to
match the precision of \wmap. On small angular scales
and for polarization measurements, however, the Planck sensitivity
is necessary and will allow significant improvements to
be made over \wmap. At these scales and sensitivities, the
wide frequency coverage of Planck will also be invaluable for
understanding foregrounds.

\subsection{Future Experiments}

The current and planned experiments will measure CMB polarization with
unprecedented sensitivity. In addition, these
experiments will teach us a great deal about the efficacy of new
techniques and detectors, about the ability to control systematics
and, of course, about astronomical foregrounds.

We are on the cusp of major advances in detector technology both in
large-format bolometer arrays and in coherent detector arrays\footnote{
see http://www.sofia.usra.edu/det\_workshop/ for a 
current review of detector
technology}. Using
integrated circuit technologies it will shortly be possible to build
correlation receivers on a single chip \citep{gaier03}; it is possible
to conceive of a CAPMAP-like experiment with hundreds of
radiometers. The ease of duplicating receivers can also be applied for
future interferometers. Interferometers, however, will be 
limited by the size of broad bandwidth correlators 
(the correlator scales as $N^2$,
where $N$ is the number of array elements).  One can conceive of
scaling the DASI and CBI design \citep{padin00} for $\sim 100$ 
receivers, but much larger correlators will require further 
advances in correlator technology.

Large format arrays for bolometers are now possible using both ``pop-up''
arrays developed at NASA/GSFC \citep[e.g.,][]{dowell03} and
monolithic arrays developed at Caltech/JPL \citep[e.g.,][]{glenn03}.
The current push in bolometer array technology is directed at micro-machined
planar arrays with superconducting transition-edge sensors (TES)
with multiplexed readouts.  Current arrays of order 1000 channels are being
developed and much larger arrays will be possible with this
technology.  The next step being actively pursued is large format
arrays of superconducting microstrip, antenna-coupled bolometer
detectors \citep[e.g.,][]{goldin03}. The antenna-coupled technology offers great flexibility;
the elements can be configured for dual polarization and support several
frequency channels.

New telescopes are required to take advantage of the
new bolometer arrays and at least three are in various
stages of planning. The South Pole Telescope (SPT)\footnote{
Collaboration led by the University of Chicago and
including the University of
California, Berkeley, Case Western Reserve University, University
of Illinois at Urbana and the Smithsonian Astrophysical Observatory} 
is
an 8-meter, precision, off-axis telescope funded by NSF-OPP to be deployed to the
South Pole station in late 2006 and equipped with a 1000-element bolometric radiometer. A large focal plane polarimeter is
planned for the SPT, but is not currently funded. 

Planning is underway for the Atacama Cosmology Telescope \citep[ACT][]{kosowsky03}, a 
6 meter off-axis telescope planned for Chile which could also be
equipped with a polarimeter. Finally, planning is also underway 
for Polarbear, a 3 meter off-axis telescope dedicated to CMB polarization
measurements with a large array (1000 to 3000 elements) 
of polarization-sensitive antenna-coupled bolometers \citep{tran03}.

The most ambitious plan on the horizon is for a dedicated CMB
polarization satellite to conduct the definitive search, i.e., foreground-limited,  for the
signature of \inflation{ary} gravitational waves in the CMB, i.e., to
measure the gravitational wave \bmode{} spectrum. It is the goal of
the Inflation Probe of NASA's {\it Beyond Einstein}
program\footnote{http://universe.gsfc.nasa.gov/}.  NASA plans to fund
of order three studies toward the eventual launch of a \$350M to
\$500M Inflation Probe.

\section{Conclusions}

The detection of CMB polarization marks the beginning of a new era in
CMB measurements and for cosmology. With the first detection of
\emode{s} and the \te{} correlation, the standard model has
a passed a critical test, and already the large-angle \te{}
correlation is pointing the way to a revised understanding of the
reionization history of the universe.

Polarization measurements are now at a turning point reached by
temperature anisotropy measurements a decade ago, when they were first
detected by COBE and FIRS \citep{smoot92,ganga93}; as was the case
with the temperature measurements, we can count on rapid progress in
the characterization of the CMB polarization. The \emode{} spectrum
should be well measured by ground-based and balloon-borne experiments,
and the {\it Planck Surveyor} satellite over the next several years.
A few experiments now underway should already be capable of measuring the
gravitationally lensed \bmode{} spectrum, and possibly even the
gravitational wave \bmode{} spectrum.

It is unreasonable, however,  to expect that increased instrument sensitivity alone
will allow the detection of these extremely weak signals. 
At the required sensitivity levels, systematics 
will be much harder
to control and contamination from astronomical foregrounds will
be much more severe. 
The need for several independent experiments using different
techniques including ground, balloon and eventually satellite-based instruments is even more important than it was for the 
temperature CMB measurements. Continued exploratory work with new techniques
and with ever more sensitive detectors is necessary to ensure
the eventual success of NASA's Inflation Probe.

\acknowledgments

We thank the organizers and staff for a stimulating and well run workshop. 
This work is supported in part by NSF grants 0094541 and 0096913.


\end{document}